\documentclass[%
 reprint,
 amsmath,amssymb,
 aps,
]{revtex4-2}

\usepackage{graphicx}
\usepackage{dcolumn}
\usepackage{bm}
\usepackage{upgreek}
\usepackage{xcolor}
\usepackage{enumitem}
\usepackage{placeins}

\begin{document}

\title{High connectivity quantum processor nodes using single-ion-qubits in rare-earth-ion-doped crystals}

\author{Adam Kinos$^1$}
\email{adam.kinos@fysik.lth.se}
\author{Lars Rippe$^1$}
\author{Diana Serrano$^2$}
\author{Andreas Walther$^1$}
\author{Stefan Kr\"{o}ll$^1$}

\affiliation{%
 $^1$Department of Physics, Lund University, P.O. Box 118, SE-22100 Lund, Sweden
}%

\affiliation{%
 $^2$Chimie ParisTech, PSL University, CNRS, Institut de Recherche de Chimie Paris, 11, rue
Pierre et Marie Curie, 75005 Paris, France
}%

\date{\today}

\begin{abstract}
We present two protocols for constructing quantum processor nodes in randomly doped rare-earth-ion crystals and analyze their properties. By varying the doping concentration and the accessible laser tunability, the processor nodes can contain anywhere from only a few tens to almost $1000$ qubits. Furthermore, the average number of qubits each qubit can interact with, denoted by the connectivity, can be partly tailored to lie between just a few and roughly one hundred. We also study how a limited tunability of the laser affects the results, and conclude that a tuning range of $100$ GHz limits the results to roughly $100$ qubits with around $50$ connections per qubit on average. In order to construct an even larger processor, the vision is that several of these quantum processor nodes should be connected to each other in a multi-node architecture via, e.g., optical interfaces or flying qubits in the form of light. Our results are encouraging for establishing the rare-earth-ion-based systems as a quantum computing platform with strong potential and can serve to focus the efforts within the field.
\end{abstract}

\maketitle

\section{\label{sec:intro}Introduction}
In the field of quantum information, rare-earth-ion-doped crystals have been used in quantum memories \cite{Nilsson2005a, Kraus2006, Hetet2008, Riedmatten2008, Afzelius2010, Hedges2010, Beavan2012a, Sabooni2013a, Dajczgewand2014, Guendogan2015, Jobez2015, Schraft2016}, conversion between optical and microwave signals \cite{OBrien2014, Williamson2014, FernandezGonzalvo2019}, and quantum computing \cite{Pryde2000, Ichimura2001, Ohlsson2002, Nilsson2002, Wesenberg2003, Wesenberg2007, Walther2009a, Walther2015, Ahlefeldt2020, Grimm2021, Hizhnyakov2021}, the latter topic is here investigated through simulations.

In order to focus the research community on the task of constructing a rare-earth quantum computer, a roadmap has been developed \cite{Kinos2021}. There, the qubits are envisioned to be single ions, and the most promising approach uses dedicated readout ions to read out the state of the qubits \cite{Wesenberg2007}. This is feasible due to the recent progress in single-ion detection \cite{Utikal2014, Xia2015, Kolesov2018, Zhong2018, Raha2020, Kindem2020}. Previous work has already indicated that single- and two-qubit gate operations with low errors \cite{Kinos2021a} are possible, even when the effects of instantaneous spectral diffusion on the single-qubit gate operations are included \cite{Kinos2021b}. The rare-earth qubits are controlled via optical pulses, where each qubit occupies a specific frequency channel, and there can be many such channels due to the large inhomogeneity in the optical transitions \cite{Konz2003}. Furthermore, the spectral addressing makes it possible to pack the ions at nanometer spacing in three dimensions. This allows for very high qubit densities. In addition, the dense packing allows for strong dipole-dipole interactions between qubits \cite{Ohlsson2002, Nilsson2002, Longdell2004a, Longdell2004, Ahlefeldt2013b}. However, the number of potential qubits in these systems and their connectivity has previously not been carefully investigated, and this is a central question for assessing the potential of rare-earth systems as a quantum computing platform. 

In this work, we examine quantum processor nodes in randomly doped rare-earth-ion crystals. There are numerous ways to select the ions that should be used as qubits, and depending on which way is used the connectivity of the processor node can vary drastically. Here we focus on two extreme protocols for selecting the qubit ions. Using the first protocol results in qubits that are densely distributed in space and have high connectivity. In contrast, using the second protocol yields qubits that are sparsely distributed and have low connectivity. We conclude that both the number of qubits and the connectivity between qubits can be partly tailored by modifying these protocols or changing physical parameters, such as doping concentration, and technological parameters, such as the available laser tunability. The quantum processor nodes in this article reach qubit numbers between a few tens and $1000$, and the average number of connections per qubit range from a few connections up to $100$. 

We now provide an overview of the sections and content of the paper. Sec. \ref{sec:system} provides all necessary information about how the processor node is constructed. Following this Sec. \ref{sec:qp_properties} investigates the number of qubits and their connectivity for processors created using different protocols and/or doping concentrations, and we also discuss the effect of a limited laser tunability. Finally, this work is concluded in Sec. \ref{sec:conc}.

\section{\label{sec:system}The construction of quantum processor nodes}

\subsection{\label{sec:material}Material overview}
\begin{figure*}
\includegraphics[width=\textwidth]{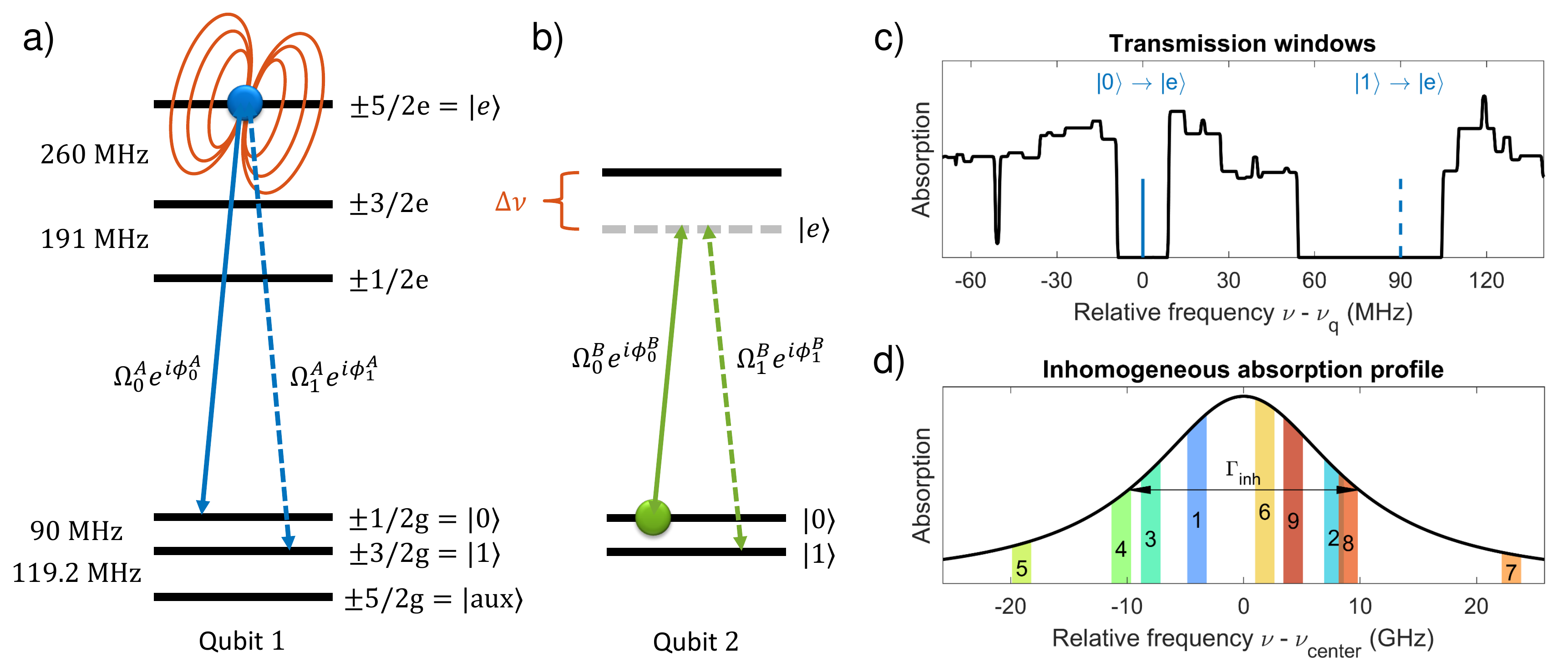}
\caption{\label{fig:Enery_levels}a) Shows the energy level structure for a qubit including the two-color pulses used to control it. The two fields driving the $|0\rangle\rightarrow|e\rangle$ and $|1\rangle\rightarrow|e\rangle$ qubit transitions are shown as solid and dashed lines, respectively. b) Shows only the relevant qubit levels for a second qubit, whose optical transitions are detuned from the first qubit. When either of the qubits are excited, the other experiences a frequency shift, $\Delta\nu$, of its optical transitions due to a dipole-dipole interaction. The strength of this interaction scales as $1/|\boldsymbol{r}|^3$, where $|\boldsymbol{r}|$ is the distance between the two qubits \cite{Jackson1998}. c) In order to minimize the effect of ISD one can empty the frequency regions surrounding the two qubit transitions using, e.g., optical pumping, such that there are no absorbing ions in those regions. Here such transmission windows with widths of roughly $18$ MHz and $50$ MHz, respectively, are shown as a function of the relative frequency $\nu - \nu_q$, where $\nu_q$ is the $|0\rangle \rightarrow |e\rangle$ transition frequency of the qubit with index $q$. d) The inhomogeneous absorption profile of the optical transitions is shown, including the full-width-at-half-maximum, $\Gamma_\text{inh}$, determined by Eq. \ref{eq:Gamma_inh}. Here we use $c_\text{total} = 1\%$ as an example. We also show a schematic representation of nine qubits with indices $1 \rightarrow 9$ and different transition frequencies $\nu_q$. The colored regions show the reserved frequency interval surrounding each qubit, and other qubits may not have their $\nu_q$ transition frequency within this interval. Note, however, that as long as the $\nu_q$ transition frequencies are outside other colored regions, those regions may themselves partly overlap. This can be seen for the qubits with indices $2$ and $8$.}
\end{figure*}

In this work we investigate a quantum processor node using site 1 $^{153}$Eu:Y$_2$SiO$_5$ (europium doped into yttrium orthosilicate), whose  relevant properties can be found in Fig. \ref{fig:Enery_levels}a. 

Single-qubit (SQ) gate operations are performed using $2$ two-color pulses resonant with the $|0\rangle\rightarrow|e\rangle$ and $|1\rangle\rightarrow|e\rangle$ qubit transitions following the procedure outlined and motivated in reference \cite{Kinos2021a}. The average SQ gate error is estimated to be $3.4\cdot 10^{-4}$ \cite{Kinos2021a}, and even if instantaneous spectral diffusion (ISD) is accounted for, the error is in the same order of magnitude for the vast majority of qubits \cite{Kinos2021b}. The effect of ISD is minimized by using spectral hole burning techniques \cite{Nilsson2002, Nilsson2004, Rippe2005, Lauritzen2012} to isolate the qubit ions in frequency space by creating semipermanent transmission windows in the inhomogeneously broadened rare-earth-ion ensemble, see Fig. \ref{fig:Enery_levels}c. How this can be done in site 1 $^{153}$Eu:Y$_2$SiO$_5$ is described further in reference \cite{Kinos2021b}.

Two-qubit (TQ) gate operations are performed using a dipole-dipole interaction which allows different qubits to interact if they are sufficiently close in space. The dipole-dipole interaction is explained in Fig. \ref{fig:Enery_levels}a-b and scales as $1/|\boldsymbol{r}|^3$ where $|\boldsymbol{r}|$ is the distance between the two qubits \cite{Jackson1998}. Two different types of TQ gates are used: the blockade gate which requires dipole-dipole shifts larger than the pulse bandwidth so that if one qubit is excited, the other qubit is shifted out of resonance with its control pulses; and the interaction gate which takes advantage of the exact dipole-dipole shift to, e.g., add a phase to a specific TQ state \cite{Kinos2021a}. When the dipole-dipole shift, $\Delta\nu$, between two qubits is sufficiently strong, $|\Delta\nu| \geq 7$ MHz, the TQ blockade gate is used, whereas the TQ interaction gate is used when the shift is less than this, $0.1$ MHz $\leq |\Delta\nu| < 7$ MHz. The average TQ gate errors are estimated to lie in the range of $5\cdot 10^{-4} \rightarrow 3\cdot 10^{-3}$ depending on the exact dipole-dipole shift, for more information see reference \cite{Kinos2021a}. Note that these error rates lie beneath the threshold for error correction using surface codes \cite{Raussendorf2007, Fowler2009} and lies at the border of the threshold for CSS codes \cite{Steane2003}. 

Although the SQ and TQ gates are not directly used in the simulations presented in this work, they do affect the properties of the processor. If one uses TQ gates that work for small shifts, and thus large separations $|\boldsymbol{r}|$ between qubits, the resulting processor has a larger connectivity compared to if TQ gates which only work for large shifts are used. 

Furthermore, the gate parameters also affect how large frequency range each qubit must reserve in order for other qubits not to cause significant additional errors, e.g., when gate operations are performed on those other qubits. Reference \cite{Kinos2021a} investigated how the control pulses of one qubit affects the state of another qubit as a function of the frequency difference between the $|0\rangle\rightarrow|e\rangle$ transitions of the two qubits, and concluded that a minimum detuning of roughly $\pm0.6$ GHz was required. However, this minimum detuning is increased even further when the effect of ISD is minimized by creating transmission windows for each qubit as described above. In Appendix \ref{app:qubit_occupation} we calculate the frequency detunings that must be reserved for each qubit to allow for the creation of such transmission windows, and find that it covers roughly $1.7$ GHz, meaning that another qubit cannot have its $|0\rangle\rightarrow|e\rangle$ transition in that interval. However, since the reserved frequency regions for different qubits can overlap as long as the $|0\rangle\rightarrow|e\rangle$ transition frequency of a qubit does not enter into another region, one can, in the best case when the qubits are packed as tightly as possible in frequency space, fit one qubit per roughly $1.7/2$ GHz $= 0.85$ GHz. A schematic view of how the qubits reserve parts of the inhomogeneous absorption profile can be seen in Fig. \ref{fig:Enery_levels}d.

\subsection{\label{sec:crystal_realization}Crystal realization}
In order to estimate the number of qubits and their connectivity for a quantum processor node we create random realizations of a crystal. Using the structure of the yttrium orthosilicate (Y$_2$SiO$_5$) host crystal \cite{Maksimov1971,Bengtsson2012, Villars2021} we randomly replace a fraction, $c_\text{total}$, of the yttrium ions with the $^{153}$Eu dopants. We assume that half of the dopants occupy site 1 which we investigate here, and thus only half of the total number of dopants can form qubits in the processor. More detailed information about the doping procedure can be found in reference \cite{Kinos2021b}. 

After this random placement we know the spatial position as well as the relative direction of the static electric dipole moment difference, $|\Delta\boldsymbol{\mu}| = 7.6 \cdot 10^{-32}$ Cm \cite{Graf1998}, for all dopants. This allows us to calculate the dipole-dipole shift, $\Delta\nu$, between any two dopants \cite{Kinos2021b, Jackson1998}:
\begin{eqnarray}\label{eq:delta_nu}
    \Delta\nu =&& \frac{k}{|\boldsymbol{r}|^3} (\Delta\boldsymbol{\mu}_A\cdot\Delta\boldsymbol{\mu}_B - 3(\Delta\boldsymbol{\mu}_A\cdot\hat{\boldsymbol{r}})(\hat{\boldsymbol{r}}\cdot\Delta\boldsymbol{\mu}_B)) \nonumber\\
    k =&& \frac{(\epsilon(0) + 2)^2}{9\epsilon(0)} \frac{1}{4\pi\varepsilon_0 h} 
\end{eqnarray}

where $\boldsymbol{r}$ is the spatial vector pointing from ion $B$ to ion $A$, and $\hat{\boldsymbol{r}}$ is the normalized spatial vector. The first term in the constant $k$ is a local field correction due to the crystal \cite{Mahan1967}, where the dielectric constant for DC fields, $\epsilon(0)$, is equal to $11$ for the case of Y$_2$SiO$_5$ \cite{Christiansson2001, Mock2018}. $\varepsilon_0$ is the vacuum permittivity and $h$ is Planck's constant. 

Following this, each dopant is randomly assigned a frequency based on a Lorentzian line shape of the inhomogeneous absorption profile as seen in Fig. \ref{fig:Enery_levels}d. For the doping concentrations studied in this work, the full-width-at-half-maximum (fwhm), $\Gamma_\text{inh}$, of the line shape is assumed to grow linearly:
\begin{equation}\label{eq:Gamma_inh}
    \Gamma_\text{inh} = \Gamma_0 + \Gamma_c \cdot c_\text{total}
\end{equation}

where $\Gamma_0 = 1.8$ GHz is a concentration independent linewidth, $\Gamma_c = 1800$ GHz \cite{Sellars2004, Bengtsson2012}, and $c_\text{total}$ specifies the total atomic doping concentration between $0$ and $1$. In a real crystal there is a slight correlation between ions having similar transition frequencies and being spatially close to each other \cite{Sellars2004}, but this is assumed to be negligible in the present work as we randomly distribute the transition frequencies of the ions. 

To account for the potentially limited tuning range of the laser used to initialize and control all qubits, any dopants that lie outside this range are now removed. In the present work we assume that the tuning range is either $100$ GHz or $1000$ GHz, and that it is centered around the inhomogeneous absorption profile. The technical challenges of reaching such high laser tunability while still maintaining the coherent and precise control required by the gate operation pulses are discussed further in, e.g., reference \cite{Kinos2021}. 

A $1000$ GHz tuning range is much larger than the maximum linewidth of the inhomogeneous absorption profile studied in this work. For example, doping concentrations of $1\%$ or $5\%$ have a linewidth of $19.8$ GHz or $91.8$ GHz, respectively. However, since we assume that the absorption profile has a Lorentzian shape \cite{Konz2003}, many qubits can still be found in the wings of the profile. Note, however, that in a real crystal the ions would eventually deviate from a Lorentzian shape and satellite lines would instead form \cite{Yamamoto1969, Sellars2004}.

Furthermore, for a $1000$ GHz tuning range the closeness between the two Eu sites, roughly $150$ GHz \cite{Equall1994}, might complicate the procedure to find qubits when constructing the quantum processor, and it might also increase the effect of ISD. However, if one allows qubits to be formed using either site, the closeness between the two sites might instead be turned into an advantage as it provides additional ions to be used as qubits. Regardless, in the present work we neglect these issues or benefits and assume that one can isolate Eu dopants from site 1 despite the overlap with dopants in site 2. 

In general, energy transfer processes can occur between any two dopants but the effect is strongly dependent on the distance between them. In Appendix \ref{app:energy_transfer} we investigate the case for site 1 $^{153}$Eu:Y$_2$SiO$_5$ and conclude that fluorescence resonant energy transfer effects are negligible. 

Finally, each crystal we create is a sphere with a $100$ nm radius. Note, however, that a real crystal does not need to be this small in order to achieve the results presented in this work. The radius was picked as a trade-off between how long the simulations take to run and how wide the qubits are spread out in space. Furthermore, since the creation of the crystal is a stochastic process we perform $100$ different crystal realizations for each protocol or parameter we want to investigate. 

\subsection{\label{sec:qp_protocol}Protocols for constructing quantum processor nodes}
\begin{figure*}
\includegraphics[width=\textwidth]{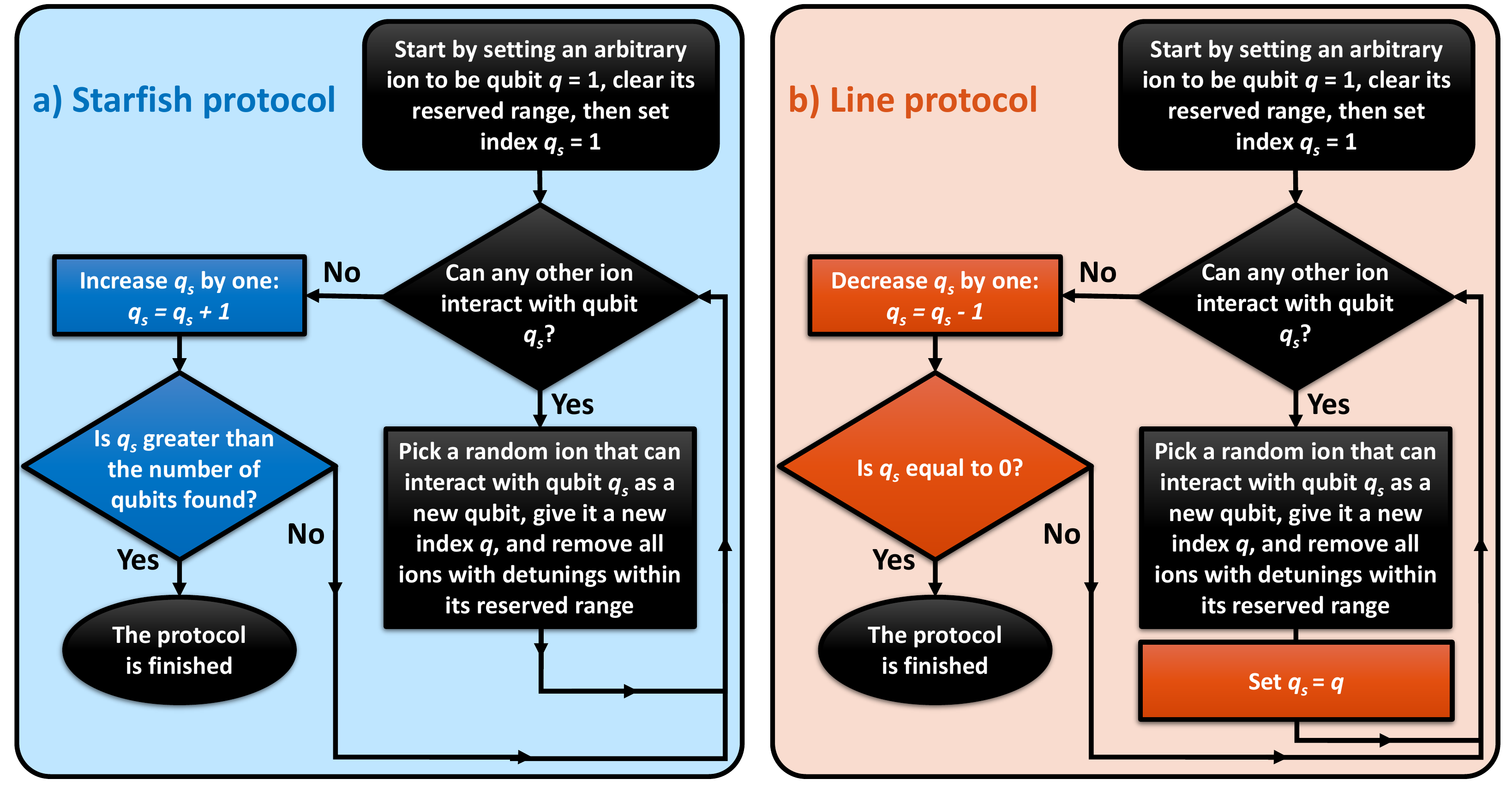}
\caption{\label{fig:flowchart}Shows the procedure to construct quantum processor nodes based on two methods: a) the starfish protocol (blue) and b) the line protocol (red), where common steps are shown in black. We give the qubits indices $q = 1,2,3...$ based on the order in which they are found, and search for new qubits connected to the qubit with index $q_s$. The starfish (line) protocol always attempts to find qubits connected to the first (last) qubit found before continuing the search around later (earlier) qubits. In an experimental setting where the qubits are read out by a dedicated co-doped readout ion, the first qubit in the protocol is set to be one ion that can turn off the fluorescence of the readout ion, for more information, see Appendix \ref{app:experimental_protocol}. \\\hspace{\columnwidth} }
\end{figure*}

\begin{figure}[!ht]
\includegraphics[width=\columnwidth]{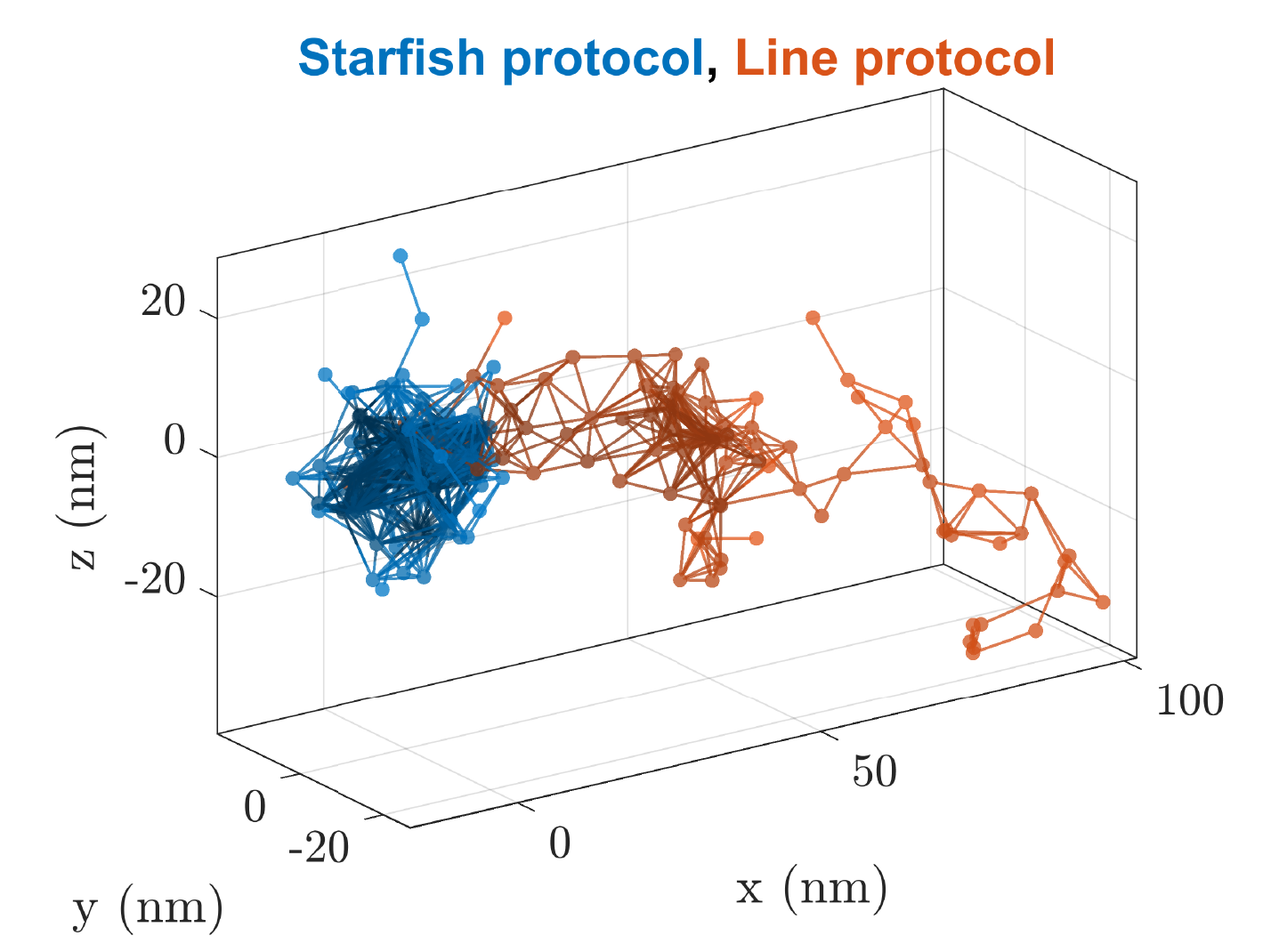}
\caption{\label{fig:qubit_spatial}Shows two examples of the spatial distribution of the qubits in the quantum processor node when constructed using the starfish protocol (blue) and the line protocol (red). The dots represents qubits and the lines represent possible TQ interactions between the connected qubits. In these simulations the doping concentration was $c_\text{total} = 1\%$ and the tuning range of the laser was $100$ GHz. \\\hspace{\columnwidth} }
\end{figure}

In this section we introduce two main protocols to find the ions that are used as qubits in the quantum processor node. We denote these as the \textit{starfish} and \textit{line} protocols. In short, the starfish protocol attempts to find new qubits surrounding the qubits found first, thus resulting in a dense spatial distribution of the qubits. On the contrary, the line protocol attempts to find new qubits surrounding the qubits found last, hence the qubits form a line of interaction that is more likely to spread out in space. Flow charts describing the two protocols are given in Fig. \ref{fig:flowchart}, and examples of how the qubits are spatially distributed can be seen in Fig. \ref{fig:qubit_spatial}. 

In both protocols we find new qubits by randomly choosing one ion that can perform a TQ gate operation with qubit $q_s$ that has an error less or equal to $3\cdot 10^{-3}$. In the simulations we find suitable candidates by checking if the dipole-dipole shift, $\Delta \nu$, they cause to qubit $q_s$ can be used to perform a TQ gate operation. If the shift is in the range of $0.1$ MHz $\leq |\Delta\nu| < 7$ MHz the interaction gate is used, whereas the blockade gate is used when $|\Delta\nu| \geq 7$ MHz. However, some shifts are not suitable as they shift the gate operation pulses into resonance with other transitions, for more information see Appendix \ref{app:TQ_shifts}. 
The ranges listed above are used since they result in TQ gate errors less or equal to $3\cdot 10^{-3}$. However, these ranges are not fixed, and trade-offs can be made. For example, by decreasing the bandwidth of the pulses used to perform a TQ blockade gate, the minimum requirement of a $7$ MHz dipole-dipole shift is decreased. This, however, comes at the cost of longer gate pulse and thus larger errors due to decay and decoherence, which currently limits the fidelity of the TQ gate operation. 

In most experimental situations the exact dipole-dipole shifts are not initially known, and how this search can be performed experimentally is therefore still an open question which might even depend on how the state of a qubit is being read out. In Appendix \ref{app:experimental_protocol} we present a possible experimental protocol which assumes that co-doped readout ions are used to read out the state of qubits via the dipole-dipole blockade mechanism \cite{Wesenberg2007, Walther2015}. 

Finally, in Sec. \ref{sec:qp_properties} we also study a modified version of the starfish protocol, namely the \textit{compact starfish}. In this protocol one starts by performing the starfish as usual, except one only looks for interactions using the TQ blockade gate. Since this gate requires larger dipole-dipole shifts than the interaction gate, the spatial structure of the compact starfish protocol is even denser than the original version. After no more qubits can be found, the starfish protocol is repeated once more, now searching for more qubits that can interact using either of the TQ gate types, i.e., 
$q_s$ is reset to $1$ and the protocol restarts by asking "Can any ion interact with qubit $q_s$?" in Fig. \ref{fig:flowchart}a, now looking for potential qubits that can perform TQ interaction gates as well. 

\begin{figure*}
\includegraphics[width=\textwidth]{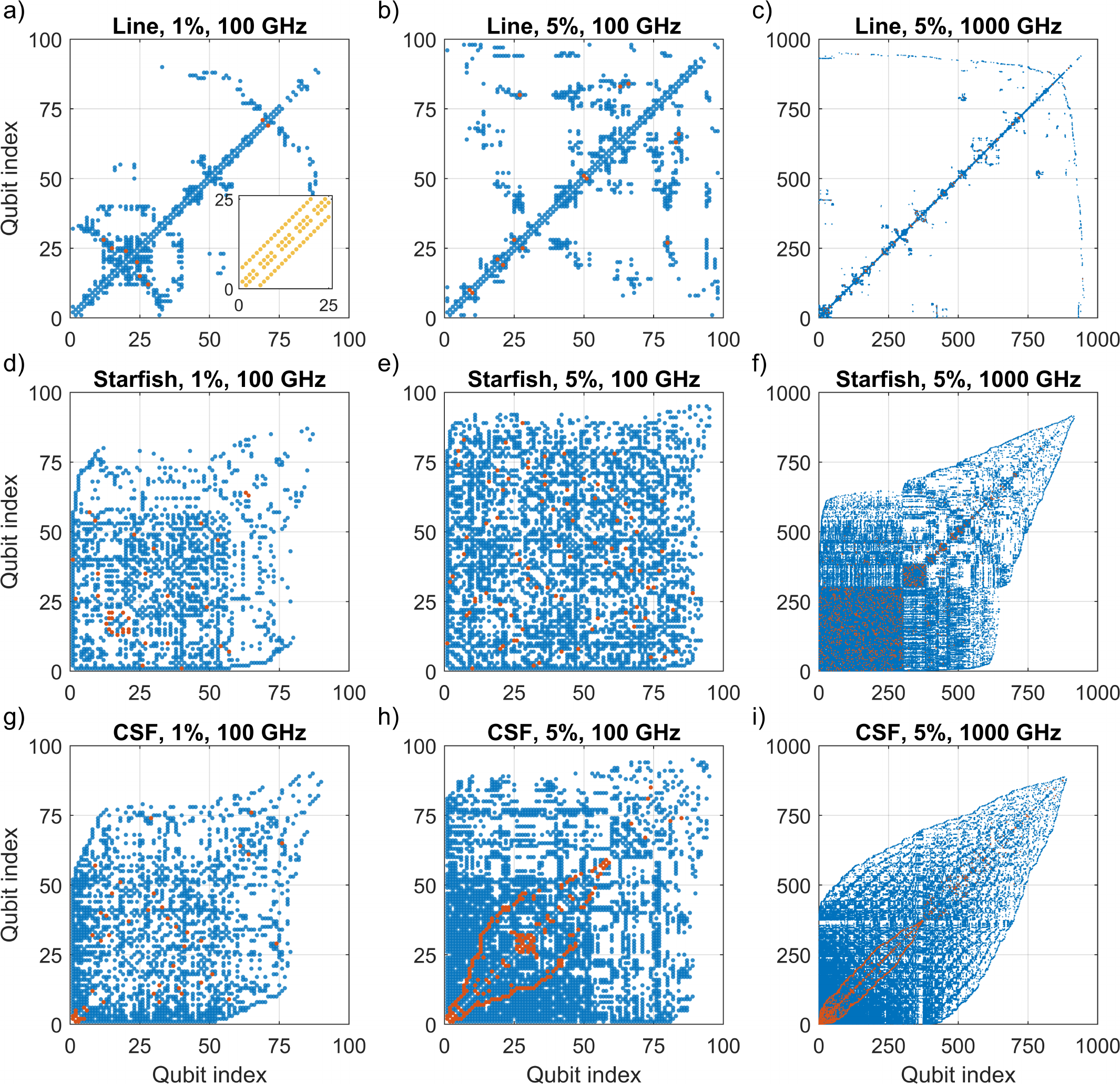}
\caption{\label{fig:qubit_connections}Shows the connectivity between different qubits, where a blue or red dot at coordinates $(i, j)$ means that the qubits with indices $i$ and $j$ can perform a TQ interaction or blockade gate, respectively. The titles for each graph explain which protocol (CSF = compact starfish), doping concentration $c_\text{total}$, and laser tuning range that was used in the construction of the processor node. These graphs show the result from one crystal realization each, but still represent typical cases for how the connectivity is distributed depending on the parameters. In order to easier compare to the connectivity of other systems, we in the inset of graph a) show the connectivity graph for the case of qubits arranged in a square grid ($5$ by $5$ grid for a total of $25$ qubits) where each qubit is connected to its immediate neighbours to the left, right, up, and down. This results in four lines, where the two central lines are cut at regular intervals since the qubits at the edges (corners) of the grid are missing one (two) neighbour(s). For a system where all qubits can directly interact with all other qubits, the connectivity graph would be completely filled. (a-c) The main diagonal line arises since the line protocol always attempts to find new qubits connected to the latest qubit found, see Fig. \ref{fig:flowchart}b. (d-f) The visual square-shaped patterns arise when many qubits form a highly connected subgroup. This occurs since the starfish protocol attempts to find as many qubits as possible connected to one qubit, see Fig. \ref{fig:flowchart}a, and all those qubits are therefore spatially close to each other. For more detailed explanations, see the main text.}
\end{figure*}

\section{\label{sec:qp_properties}Quantum processor node properties}
In this section we investigate the properties of quantum processor nodes constructed in different ways. We start by examining the connectivity for the line protocol when the doping concentration is $c_\text{total} = 1\%$ and the laser tuning range is $100$ GHz. Fig. \ref{fig:qubit_connections}a shows the connectivity graph between all qubits for this case. This looks as one could expect based on the spatial distribution seen in Fig. \ref{fig:qubit_spatial} (red line), i.e., each qubit is mainly connected to just a few other qubits in its vicinity as indicated by the diagonal line in the connectivity graph. It is, however, still possible that qubits clump together and form a slightly more connected subgroup. Furthermore, as it becomes increasingly more difficult to find new qubits connected to the last qubit found, the line protocol dictates that one should try to find qubits connected to earlier qubits, which explains why the qubits with large indices are sometimes connected to earlier qubits and therefore deviate from the main diagonal line. 

If the doping concentration increases, the average number of connections increases slightly. The reason for this is that a higher concentration on average leads to shorter distances between qubits, thus increasing the connectivity slightly. The higher concentration also makes it easier to find more qubits. A connectivity graph example for the line protocol where $c_\text{total} = 5\%$ can be seen in Fig. \ref{fig:qubit_connections}b. 

Presently, the number of qubits found is limited by the $100$ GHz tuning range of the laser. This can be understood since in the best case each qubit occupies a frequency range of $0.85$ GHz, thus setting an upper limit on the number of qubits to $\frac{100 \text{ GHz}}{0.85 \text{ GHz}} \approx 120$. Thus, after finding roughly $100$ qubits most optical frequencies within the tuning range are reserved and the probability of finding more ions that are connected to a qubit is low. However, if the tuning range is increased more qubits can be found. The results of the line protocol with $c_\text{total} = 5\%$ and a $1000$ GHz tuning range is shown in Fig. \ref{fig:qubit_connections}c. As can be seen, the overall structure of the connectivity graph remains similar regardless of tuning range and concentration. 

We now turn our investigation toward the starfish protocol. However, this time we begin by examining the case where $c_\text{total} = 5\%$ and the laser tuning range is $1000$ GHz. In the starfish protocol we find all qubits connected to qubit $1$ first. In our case, shown in Fig. \ref{fig:qubit_connections}f, the first qubit can interact with the first roughly $300$ qubits. An interaction is only possible if the dipole-dipole shift is at least $\pm100$ kHz, see Appendix \ref{app:TQ_shifts}, which corresponds to a maximum distance between the qubits in the order of $d = 10$ nm. For simplicity we can assume that an individual qubit can interact with all qubits within a sphere of radius $d$. Thus, the first $300$ qubits all lie within the sphere centered at the first qubit. Furthermore, the sphere for any one of those $300$ qubits have a large overlap with the sphere centered at the first qubit. Therefore, all these qubits form a highly connected subgroup of the processor. This is exactly what we see in Fig. \ref{fig:qubit_connections}f, where the connectivity graph of these qubits form a highly filled box. 

After all qubits connected to qubit $1$ has been found in the starfish protocol, the procedure continues with finding qubits connected to qubit $2$, then qubit $3$, and so on. This creates secondary boxes along the diagonal which are formed due to the same reason as explained above, except the spheres are now centered around qubit $2$, qubit $3$, and so on. 

\begin{figure*}
\includegraphics[width=\textwidth]{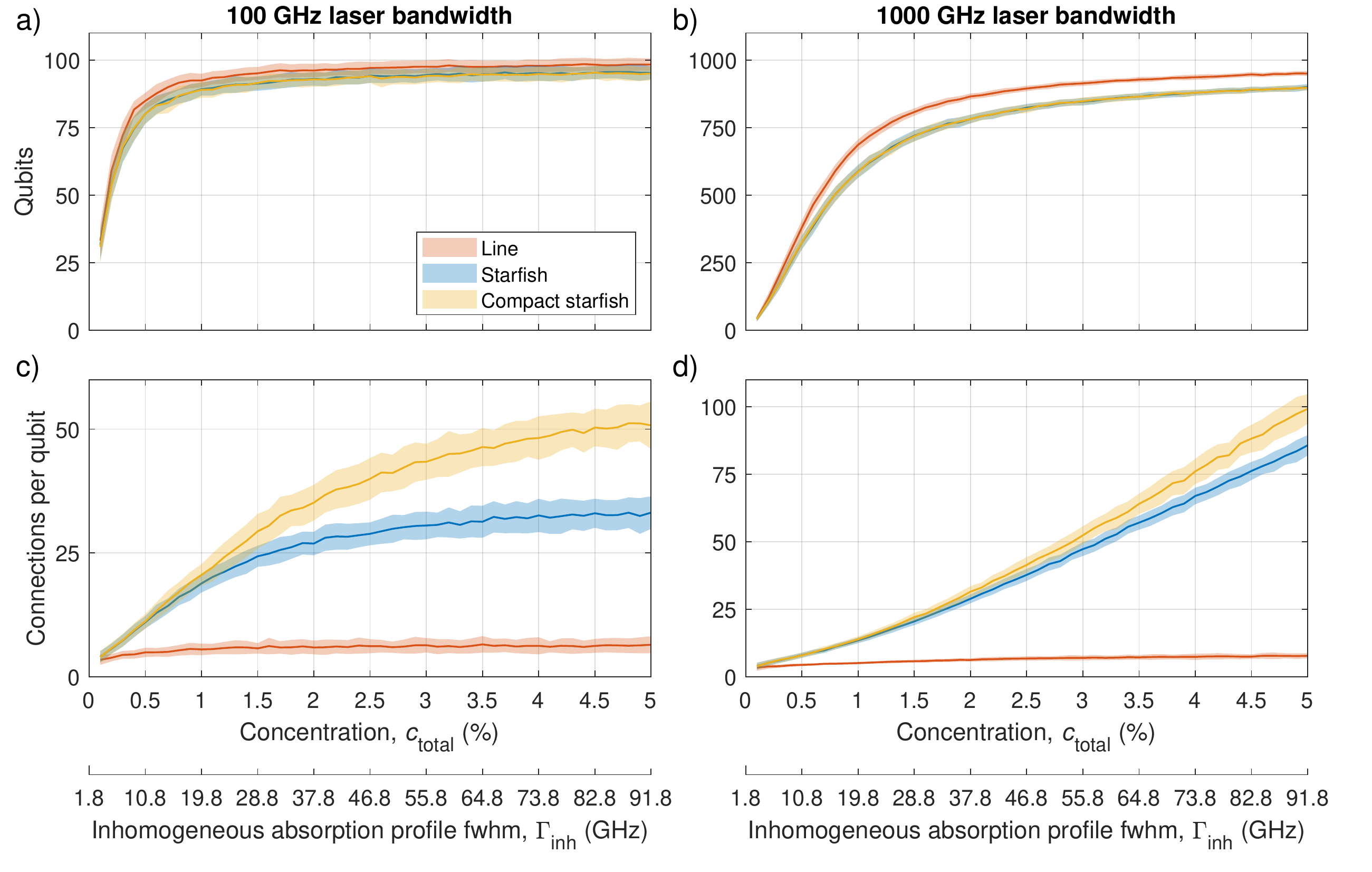}
\caption{\label{fig:qubit_chain}Shows the a-b) average number of qubits and c-d) average number of connections per qubits in the processor node, as a function of the total doping concentration, $c_\text{total}$, and the corresponding fwhm of the inhomogeneous absorption profile, $\Gamma_\text{inh}$, which is calculated from Eq. \ref{eq:Gamma_inh}. Each data set is constructed from investigating $100$ different crystal realizations, where the solid lines show the average result and the transparent regions show one standard deviation. In a) and c) the laser tuning range is limited to $100$ GHz, whereas the range is $1000$ GHz in b) and d). In all cases the tuning range is centered on the inhomogeneous absorption profile. The red (blue/yellow) data uses the line (starfish/compact starfish) protocol to construct the processor node. }
\end{figure*}

Once more, the number of qubits in the processor becomes limited to roughly $100$ if the tuning range of the laser is lowered to $100$ GHz, as seen in Fig. \ref{fig:qubit_connections}e. Finally, if the concentration decreases to $1\%$ the average distance between ions increases, and hence fewer qubits are found within the compact spatial distribution of the starfish protocol. It is therefore harder to find more qubits connected to qubit $1$, and thus the size of the connectivity box shrinks as seen when comparing the results of Fig. \ref{fig:qubit_connections}d and \ref{fig:qubit_connections}e. 

Fig. \ref{fig:qubit_connections}g-i show the results of the compact starfish protocol, which is discussed more at the end of this section. 

The connectivity of the starfish and compact starfish protocols are much greater than the line protocol. There is, however, a potential benefit in using the line protocol. The limiting factor on the number of qubits in the processor node is directly related to the number of available frequency channels, i.e., the ratio between the tuning range of the laser and how large frequency interval each qubit reserves. However, if the laser used to control the qubits has a small enough focus size compared to the spatial extent of the qubits, one can translate the laser focus so that the light no longer interacts with some of the qubits. The reserved frequency range of those qubits can therefore be reused since qubits only interact with the laser if both the correct frequency and the correct spatial location of the laser focus is used. This opens up possibilities for future protocols which scale beyond the constrains of a single processor node, e.g., by using thin film crystals and a translatable laser focus. 

Finally, a common feature to all connectivity graphs is that the number of TQ interaction gates (blue dots) heavily outnumber the number of TQ blockade gates (red dots). This occurs since TQ interaction gates work for smaller dipole-dipole shifts, and thus larger distances compared to the TQ blockade gates. 

Fig. \ref{fig:qubit_connections} examined how the connectivity of processors might look in a few different cases. Now, we instead investigate the number of qubits and connections per qubit in the processor when averaging the data of $100$ different crystal realizations. The results for all protocols is shown in Fig. \ref{fig:qubit_chain} as a function of the doping concentration.

As can be seen in Fig. \ref{fig:qubit_chain}a, the line protocol (red) results on average in slightly more qubits compared to the starfish protocol (blue). Furthermore, as the doping concentration increases from $c_\text{total} = 0.1\%$ to $1\%$ the number of qubits in the processor grows from roughly $30$ to $90$. However, as the concentration continues to grow the number of qubits remains fairly constant at around $100$. The reason for this saturation is the limited tuning range of $100$ GHz as explained previously. If the tuning range is increased to $1000$ GHz up to ten times more qubits can be found, as is seen in Fig. \ref{fig:qubit_chain}b. 

Since there are more ions with resonance frequencies close to the center of the inhomogeneous absorption profile, see Fig. \ref{fig:Enery_levels}d, it is more likely that those frequency channels are reserved first. However, each time we select another qubit we at least reserve a frequency range of $0.85$ GHz to that qubit. Thus, to find more qubits we must find ions in the outskirts of the inhomogeneous absorption profile. The initial increase in the number of qubits when the concentration grows from a low value can therefore be explained by an increased probability of finding ions with large detunings from the inhomogeneous absorption profile that also lie sufficiently close in space to an already existing qubit in the processor. 

We now continue our investigation by examining the average number of connections each qubit has, i.e., how many other qubits can a qubit interact with using the TQ gate operations. This is shown in Fig. \ref{fig:qubit_chain}c-d for laser tuning ranges of $100$ GHz and $1000$ GHz, respectively. Here the difference between the line and starfish protocols is more evident, with the starfish protocol having many more connections per qubit since the ions form a more compact spatial formation, as seen in Fig. \ref{fig:qubit_spatial}. Furthermore, even though the number of qubits in the processor saturates at a fairly low concentration, the average number of connections in the starfish protocol continues to grow for much longer. As explained previously this is also visualized by the increased size of the box in the connectivity graph as shown in Fig. \ref{fig:qubit_connections}e. 

When increasing the doping concentration two effects occur simultaneously: the total number of dopants increases; and the inhomogeneous absorption profile is broadened as determined by Eq. \ref{eq:Gamma_inh}. In Appendix \ref{app:fixed_conc_Gamma} we investigate how these effects individually affect the number of qubits and the connectivity of the processor node. The conclusions are that increasing the width of the inhomogeneous absorption profile can result in more qubits and more connections, but mainly if the width was originally much less than the tuning range of the laser. The width can, e.g., be increased by co-doping the crystal with another rare-earth ion \cite{Bottger2008}. Furthermore, increasing the number of dopants without broadening the inhomogeneous absorption profile always leads to more qubits and connections. However, neither effect impacts the number of qubits if it is already saturated by the limited tuning range of the laser.

Finally, the line and starfish protocols presented in this work can be modified or combined in order to partly obtain control over how the final connectivity graph should look. One example of this is the compact starfish protocol which at first only allows interactions via TQ blockade gates as explained in Sec. \ref{sec:qp_protocol}. Thanks to its denser spatial structure even when compared to the original starfish protocol, this modified protocol achieves more connections per qubit on average, as can be seen in Fig. \ref{fig:qubit_chain}c-d. For completeness we also provide a few examples of the connectivity graphs for this protocol, see Fig. \ref{fig:qubit_connections}g-i.

\section{\label{sec:conc}Conclusion}
In this work we have presented two main protocols, as well as a modified protocol, to construct quantum processor nodes using randomly doped rare-earth-ion crystals. The compact starfish protocol generates the densest spatial distribution of the qubits and thus have more connections per qubit on average. For a laser with a tuning range of $100$ GHz this protocol achieves almost $100$ qubits with roughly $50$ connections per qubit. Furthermore, if the tuning range is increased to $1000$ GHz these properties are increased to almost $1000$ qubits with roughly $100$ connections per qubit on average. 

We also conclude that increasing the width of the inhomogeneous absorption profile can be beneficial as long as the width is less than the tuning range of the laser. Furthermore, increasing the doping concentration up to at least $5\%$ was always beneficial as it increased the number of connections between qubits. 

These results assume gate operation parameters that have numerically been shown to have low gate errors \cite{Kinos2021a, Kinos2021b}. Thus, the processor nodes with the properties listed above should be ready for error correction. Our vision is then that several nodes can be connected to each other in a multi-node architecture in order to further scale up the number of qubits, e.g., via optical interfaces and communication by flying qubits in the form of light \cite{Debnath2021, Kinos2021}. 

Finally, the line protocol has fewer connections on average, but its qubits are spread out more in space. This opens up other possibilities for future scaling beyond the limits of a single processor node. For example, a translatable laser that is focused onto a small spot on a thin film could allow for the reuse of qubit frequency channels as each qubit is only interacting with the laser light if the correct frequency and the correct spatial position of the laser focus is used.

\begin{acknowledgments}
This research was supported by Swedish Research Council (no. 2016-05121, no. 2015-03989, no. 2016-04375, and 2019-04949), the Knut and Alice Wallenberg Foundation (KAW 2016.0081), the Wallenberg Center for Quantum Technology (WACQT) funded by The Knut and Alice Wallenberg Foundation (KAW 2017.0449), and the European Union FETFLAG program, Grant No.820391 (SQUARE).
\end{acknowledgments}

\appendix

\begin{figure*}
\includegraphics[width=\textwidth]{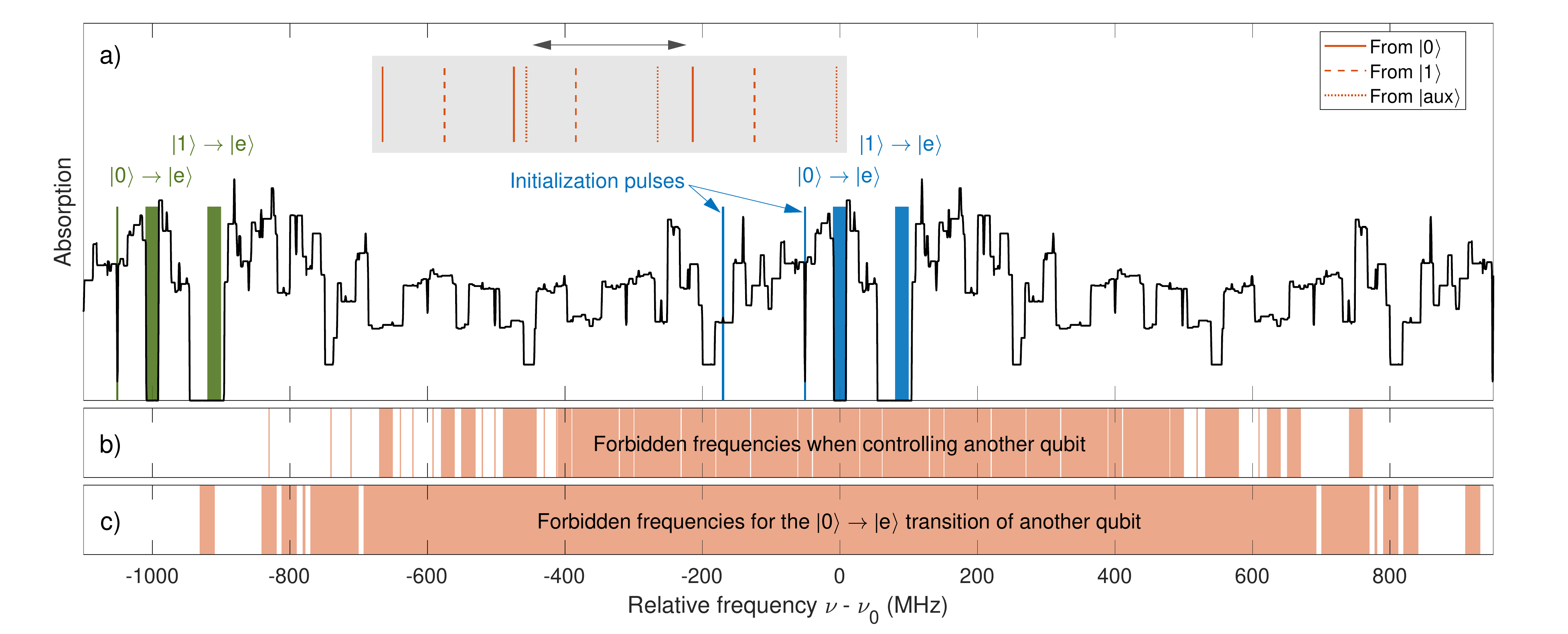}
\caption{\label{fig:cleaning}a) Shows the absorption spectrum as a function of the relative frequency $\nu - \nu_0$, where $\nu_0$ is the $|0\rangle \rightarrow |e\rangle$ transition frequency of qubit $0$. The blue regions indicate where we might send in light in order to control or reinitialize qubit $0$. The wider regions are $20$ MHz broad and covers the frequency bandwidth of the gate operation pulses, whereas the narrower regions (indicated by the blue arrows) are only $2$ MHz broad and are only used to initialize the qubit from $|\text{aux}\rangle \rightarrow |\text{3/2e}\rangle \rightarrow |1\rangle$. In order to minimize instantaneous spectral diffusion (ISD), the pulses controlling a second qubit, shown in green, should not contain frequencies which could optically pump non-qubit ions back into any of the frequencies used to control the blue qubit. We now calculate which those frequencies are. In the grey box we show the nine transition frequencies of a non-qubit ion (red), where solid/dashed/dotted lines show transitions from the $|0\rangle$/$|1\rangle$/$|\text{aux}\rangle$ ground state, respectively (the energy level structure used to calculate the transition frequencies is shown in Fig. \ref{fig:Enery_levels}a). We translate the grey box in frequency and anytime a transition frequency overlaps with one of the blue regions we mark the frequencies of the red transitions from the two other ground states as forbidden. For example, the highest frequency transition in the grey box (rightmost dotted red line), which indicate a transition from the $|\text{aux}\rangle$ state, overlaps with a blue region. Therefore, we are not allowed to use any frequency where a transition from either $|0\rangle$ (solid red lines) or $|1\rangle$ (dashed red lines) exist, since that might optically pump the non-qubit ion into $|\text{aux}\rangle$ from where it may cause significant ISD errors when an operation is performed on the blue qubit. b) Shows the frequencies that are not allowed when controlling another qubit, as obtained by the method described above. Hence, the green regions, which are the frequencies used to control another qubit, are not allowed to overlap with these frequencies. Since it is known which four frequency intervals we must use to control the green qubit, it is possible to calculate which frequencies are forbidden for the $|0\rangle \rightarrow |e\rangle$ transition of the green qubit. These frequencies are shown in graph c). In conclusion, the $|0\rangle \rightarrow |e\rangle$ transition of another qubit cannot have a frequency that overlaps with the spectrum shown in graph c).}
\end{figure*}

\section{\label{app:qubit_occupation}Calculating the reserved frequencies for a qubit}
In this section we obtain which frequencies each qubit must reserve so that transmission windows can be created for all qubits. Note that this requirement is only necessary if instantaneous spectral diffusion (ISD) is deemed to be a problem and therefore should be minimized, which we assume for the present work. The procedure to calculate these reserved frequencies is shown in Fig. \ref{fig:cleaning}. The conclusion is that a single qubit, shown in blue, reserves the frequency intervals shown in graph c). In total these intervals cover roughly $1.7$ GHz. However, the graph only shows what frequencies are forbidden for the $|0\rangle \rightarrow |e\rangle$ transition of another qubit. Therefore, the reserved frequencies of different qubits may partly overlap. In the best case roughly half of the frequency intervals can overlap, which corresponds to a qubit reserving $1.7/2$ GHz $= 0.85$ GHz, which cannot be used by another qubit.

\begin{figure*}
\includegraphics[width=\textwidth]{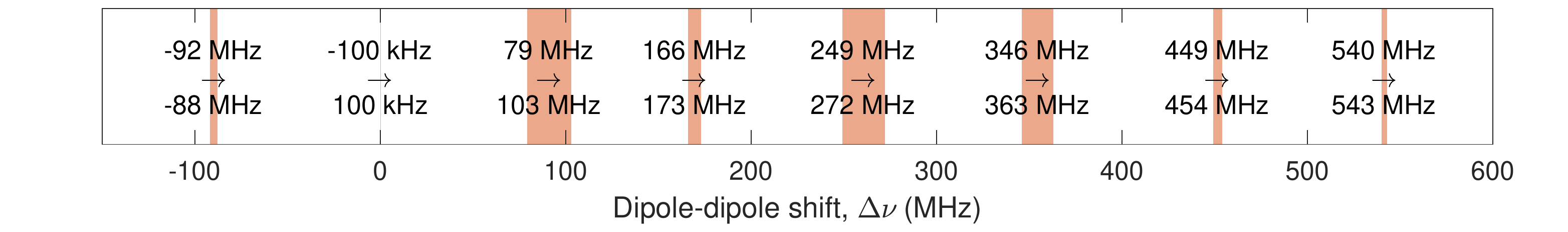}
\caption{\label{fig:dipole_ranges}In order for the TQ gate errors of the blockade and interaction gates with gate parameters found in reference \cite{Kinos2021a} to be lower or equal to $3\cdot 10^{-3}$, the dipole-dipole shift, $\Delta\nu$, between two interacting qubits cannot lie in any of the intervals listed above. }
\end{figure*}

\section{\label{app:energy_transfer}Energy transfer between Eu dopants in Y$_2$SiO$_5$}
In general, the fluorescence resonant energy transfer (FRET) rate for a dipole-dipole interaction between two dopants is given by \cite{Forster1948, Dexter1953, Serrano2014}:
\begin{equation}\label{eq:energy_transfer}
    A_{ET} = \frac{C_{dd}}{|\boldsymbol{r}|^6}
\end{equation}

where $|\boldsymbol{r}|$ is the distance between the dopants, and $C_{dd}$ is the transfer microparameter for a dipole-dipole interaction and is calculated using \cite{Caird1991}:
\begin{equation}\label{eq:Cdd}
    C_{dd} = \frac{3c}{8\pi^4n^2} \int \sigma_{\text{Em,D}}(\lambda) \sigma_{\text{Abs,A}}(\lambda) d\lambda
\end{equation}

where $\sigma_{\text{Em,D}}$ and $\sigma_{\text{Abs,A}}$ are the donor emission and acceptor absorption cross-sections, respectively, $c$ is the speed of light in vacuum, and $n$ is the index of refraction of the host, which is $1.78$ for Y$_2$SiO$_5$ \cite{Beach1990}. 

Averaged absorption cross-sections for the ${^7}F_0 \rightarrow {^5}D_0$ transition for both Eu sites were obtained by measuring the polarized absorption spectra along all three principal axes, where some parts of the polarization-averaged spectra are published in reference \cite{Serrano2014}. The data was obtained at $10$ K using a Varian Cary 6000i spectrophotometer with a $0.1$ nm resolution. 

The ${^5}D_0 \rightarrow {^7}F_0$ emission cross-sections were estimated based on the data published in \cite{Huang2014}. The data was recorded using an Oriel spectrometer (MS125) with an Andor CCD camera as detector, the sample was kept at $4$ K, and the excitation was performed at $370$ nm. The spectral resolution is estimated to be around $0.5$ nm. Both the emission and absorption measurements were recorded in a $0.1 \%$ Eu:Y$_2$SiO$_5$ crystal.

From these results the $C_{dd}$ parameter could be calculated: $C_{dd} \approx 0.43$ nm$^6$/s for site 1 (and $\approx0.24$ nm$^6$/s for site 2, but this value is not relevant for this work). The value provides a rough order of magnitude, and even in the worst case we estimate that the value does not grow by more than a factor of $6$. 

Finally, energy transfer effects are only significant if $A_{ET}$ is of similar size compared to the radiative decay set by $A_{rad} = 1/T_1$, where $T_1=1.9$ ms for the optical transition \cite{Equall1994}, i.e., $A_{ET} \approx A_{rad}$. This occur when the distance between the two dopants is $|\boldsymbol{r}| \approx (C_{dd}T_1)^{1/6} \approx 0.3$ nm. The shortest distance between two dopants in Y$_2$SiO$_5$ is roughly $0.35$ nm, and since most dopants are much farther apart than this, energy transfer effects are assumed to be negligible. 

\section{\label{app:TQ_shifts}Unusable dipole-dipole shifts for TQ gate operations}
When performing TQ blockade gate operations, if the dipole-dipole shift, $\Delta\nu$, is such that it moves another transition from either $|0\rangle$ or $|1\rangle$ into resonance with one of the gate operation pulses, the TQ gate does not work, see reference \cite{Kinos2021a} for more details. Furthermore, when performing TQ interaction gate operations, if $|\Delta\nu| < 100$ kHz the operation takes too long and thus the error is too high. Therefore, the simulations in this work does not look for interactions between qubits if the dipole-dipole shifts is in one of the ranges shown in Fig. \ref{fig:dipole_ranges}. For any other dipole-dipole shift the TQ gate error is less or equal to $3\cdot 10^{-3}$.

\section{\label{app:experimental_protocol}An experimental protocol for constructing quantum processor nodes}
In this section we present a protocol for constructing a quantum processor node in an experimental setting. The method assumes that co-doped readout ions are used to read out the state of qubits via the dipole-dipole blockade mechanism \cite{Wesenberg2007, Walther2015}. 

After locating a readout ion with strong fluorescence, the next step is to find a qubit ion that can turn this fluorescence off via the dipole-dipole blockade mechanism. This can be done by exciting various frequency intervals in the inhomogeneous absorption profile of the qubit ions and see if the readout fluorescence disappears. A suitable frequency width of the intervals might be between $1$ MHz and $10$ MHz, depending on how likely it is to find a qubit ion that can block the readout. If the fluorescence disappears it means that at least one qubit ion with a transition within the frequency interval has a sufficiently strong interaction with the readout ion. 

The next step is to work out exactly which transition the potential qubit ion was excited on. By assuming that the ion was originally excited on a specific transition one can use optical pumping techniques \cite{Nilsson2002, Nilsson2004, Rippe2005, Lauritzen2012} to force the ion into a specific frequency interval if the assumed transition was correct. By then repeating the excitation within that frequency interval one can determine if the assumed transition was correct or not depending on if the fluorescence disappears or not. If not, one repeats the procedure by assuming that the qubit ion was originally excited on one of the other optical transitions. 

After finding the correct transition, one should narrow down the uncertainty in its transition frequency by exciting ever smaller frequency intervals, and then benchmark a SQ gate operation to optimize both the optical frequencies and the Rabi frequencies of the two qubit transitions. Both the search procedure and the narrowing of uncertainty in the transition frequencies might be sped up by using binary search methods. 

The protocol proceeds by finding a second qubit that can interact with the first qubit. The method is similar to that described above, except we do not search the frequency intervals that lie within the reserved intervals for the first qubit as described in Appendix \ref{app:qubit_occupation}, and we now look for a new qubit ion that can block a NOT operation on the first qubit. In more details, we initialize the first qubit in state $|1\rangle$, excite a frequency interval as above, attempt to perform a NOT operation on the first qubit, and lastly after the ions in the excited frequency interval have decayed to the ground state again we attempt to excite the first qubit from state $|1\rangle$ and see if the readout fluorescence is turned off or not. If it turns off, we know that at least one ion within the frequency interval interacts strongly enough with the first qubit to prevent the NOT operation, since if the NOT operation works the first qubit moves to state $|0\rangle$ and is therefore never excited, thus the readout fluorescence is never turned off. The reason we wait for the ions in the frequency interval to decay from the excited state is that they otherwise may be able to directly block the readout ion, which is not what we are searching for in this protocol. Similarly as above we then narrow the uncertainty in the optical frequency of the second qubit before we benchmark its properties, including TQ gate operations with the first qubit. 

In order for this protocol to find qubits that interact even when the dipole-dipole shift between two qubits is small, $0.1$ MHz $\leq |\Delta \nu| < 7$ MHz, the NOT operations performed above must have narrower bandwidth and thus longer gate durations than the normal SQ gate operations. This is, however, not a problem since the increased error on the NOT operation only comes in when we are searching for new qubits and does therefore not affect the performance of running algorithms on the processor node once it has been constructed. 

The procedure described above can be generalized to find more qubits until no more frequency intervals are available, or until no more qubits can be found. The final step of constructing the quantum processor node is to benchmark all SQ and TQ gates, including TQ gates between qubits that never interacted during the search protocol, and determine which qubits that directly can be read out by the readout ion. Finally, note that even if the construction of the quantum processor node would require many pulses and take a long time, the search only has to be done once. 

\begin{figure*}
\includegraphics[width=\textwidth]{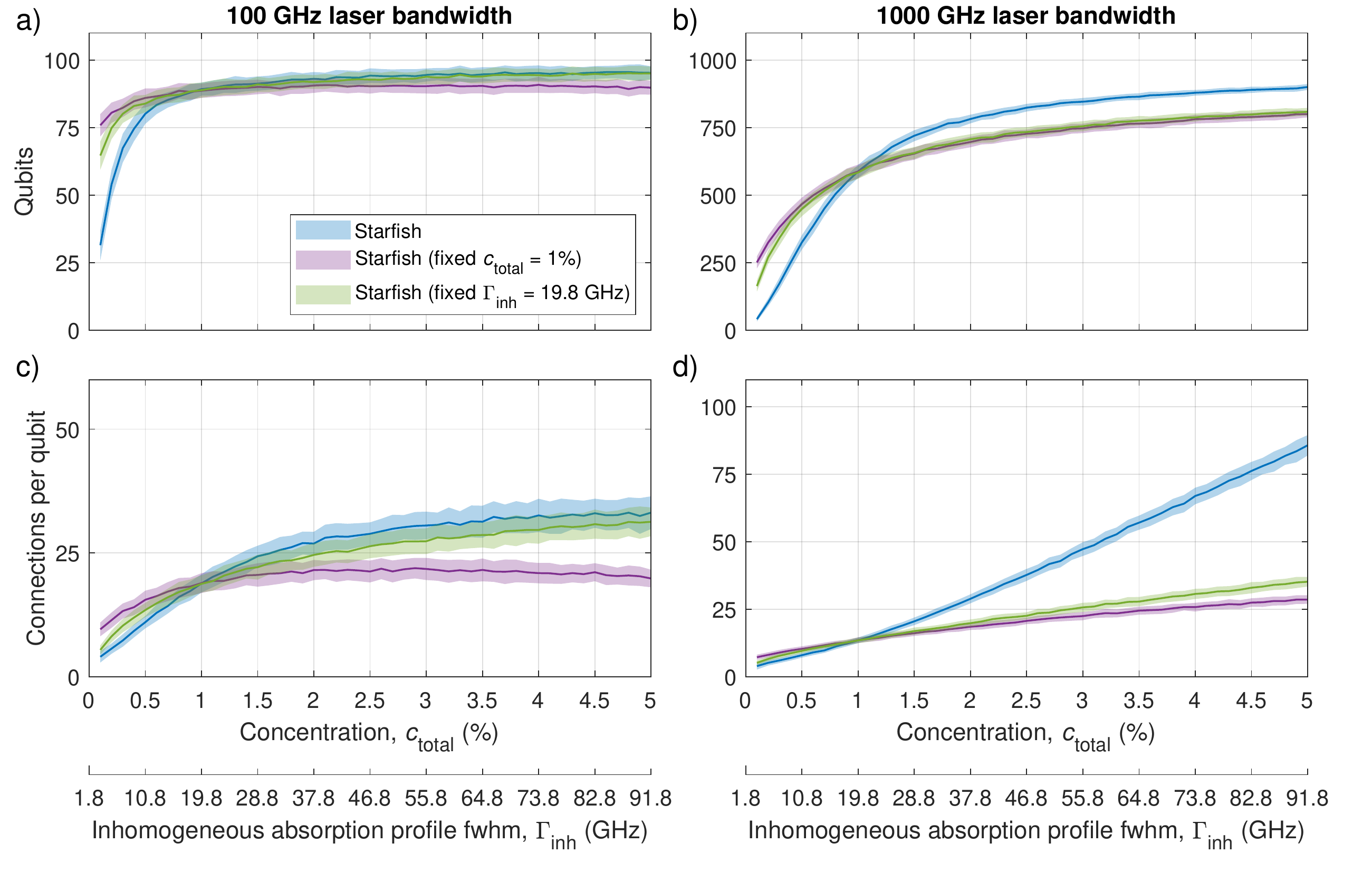}
\caption{\label{fig:qubit_chain_fixed}Shows similar graphs as Fig. \ref{fig:qubit_chain}, but we now only study the starfish protocol. The purple data keeps the doping concentration fixed at $c_\text{total} = 1\%$ and only varies the fwhm of the inhomogeneous absorption profile. For this data one should thus only use the $\Gamma_\text{inh}$ horizontal axes. Note that experimentally it might be difficult to achieve the purple data for $\Gamma_\text{inh} < 19.8$ GHz, since this is the minimum fwhm as set by Eq. \ref{eq:Gamma_inh} for a doping concentration of $c_\text{total} = 1\%$. Lastly, the green data keeps the inhomogeneous absorption width fixed at $19.8$ GHz and only varies the concentration, and hence only the concentration axes should be used for this data. Note that it might experimentally be difficult to achieve the green data for concentrations $c_\text{total} > 1\%$ since increasing the concentration normally increases the inhomogeneous broadening according to Eq. \ref{eq:Gamma_inh}.}
\end{figure*}

\section{\label{app:fixed_conc_Gamma}Investigating the impact of increasing the doping concentration}
This section examines why the number of qubits and their connectivity changes when the doping concentration changes. Note that two effects occur simultaneously when the concentration is increased: the total number of dopants increases; and the inhomogeneous absorption profile is broadened as determined by Eq. \ref{eq:Gamma_inh}. To investigate these effects we perform new simulations using the starfish protocol. First, the doping concentration is kept fixed at $c_\text{total} = 1\%$, but the inhomogeneous absorption profile is still being broadened. Experimentally this can be achieved by, e.g., co-doping the crystal with another rare-earth species \cite{Bottger2008}. Second, the concentration changes but the fwhm of the inhomogeneous absorption profile, $\Gamma_\text{inh}$, is kept fixed at $19.8$ GHz, which is the width the profile has for a $1\%$ doping concentration in Eq. \ref{eq:Gamma_inh}. This second case is difficult to realize experimentally since increasing the doping concentration normally increases the inhomogeneous broadening. Therefore, this case should primarily be used as a tool to investigate the effect of purely increasing the number of dopants in the crystal. The simulation results are shown in Fig. \ref{fig:qubit_chain_fixed}. 

We start by examining the fixed doping concentration case (purple data) for the $100$ GHz laser tuning range, see Fig. \ref{fig:qubit_chain_fixed}a and c. Here the number of qubits in the processor has already saturated for the fixed $1\%$ doping concentration, and neither the number of qubits nor the number of connections increase as the inhomogeneous width is increased above $19.8$ GHz. In other words, the effect of increasing the inhomogeneous width is negligible for a $1\%$ doping concentration and a $100$ GHz tuning range. However, for the $1000$ GHz tuning range case, both the number of qubits and connections increase as the inhomogeneous absorption profile is broadened, as is seen in Fig. \ref{fig:qubit_chain_fixed}b and d. 

For the fixed inhomogeneous width (green data), the number of qubits exhibit similar behavior as the fixed concentration case. However, even in the $100$ GHz case the number of connections now continue to grow as the concentration is increased beyond $1\%$. 

In order to explain these effects it is useful to examine how many qubits are connected to qubit $1$. When the concentration is fixed and only $\Gamma_\text{inh}$ increases, the number of ions, and thus also the number of qubits, in the center of the inhomogeneous absorption profile decreases, whereas the numbers increase in the outskirts of the profile. For the $100$ GHz tuning range case these two effects compensate each other and the total number of qubits connected to qubit $1$ remains fairly constant even when $\Gamma_\text{inh}$ changes. Therefore, the size of the connectivity box shown in Fig. \ref{fig:qubit_connections}d remains roughly the same size, which explains why the number of connections per qubit does not increase when $\Gamma_\text{inh}$ increases in the $100$ GHz tuning range case. 

When we instead have a fixed width and only the doping concentration increases, the number of ions at all frequencies increase. Therefore, the probability of finding new qubits at any transition frequency increases. Hence, the number of qubits connected to qubit $1$ increases, the connectivity box increases, and the average number of connections per qubit increases. 

In the $1000$ GHz tuning range cases the same explanations are valid, except that now when the concentration is fixed and only $\Gamma_\text{inh}$ increases the additional number of qubits found in the outskirts of the inhomogeneous absorption profile outweighs the decrease in the number of qubits in the center of the profile. This is why the number of qubits and the number of connections in the $1000$ GHz fixed concentration case continues to grow. 

\bibliography{Ref_lib}

\end{document}